\documentclass{ragtime}
\usepackage{times}
\usepackage{bibentry}
\usepackage{color}  
\title[On  the metric bundles of axially symmetric spacetimes]%
 {On  the metric bundles of axially symmetric spacetimes}

\author[D. Pugliese and H. Quevedo]
 {Daniela Pugliese\at[]{1,a} 
 and Hernando Quevedo\at[]{2}\\
 \ins{1}Institute of Physics and Research Centre of Theoretical Physics and Astrophysics,\splitins[1]
 Faculty of Philosophy \& Science,
 Silesian University in Opava,\splitins[1]
 Bezru\v{c}ovo n\'{a}m\v{e}st\'{i} 13, CZ-74601 Opava, Czech Republic
 \\
 \ins{2}
 Dipartimento di Fisica, Universit\`a di Roma ``La Sapienza", I-00185 Roma,\splitins[1] Italy Instituto de Ciencias Nucleares, Universidad Nacional Aut\'onoma de M\'exico,\splitins[1]  AP 70543, M\'exico, DF 04510, Mexico
 \ins{a}\Email{d.pugliese.physics@gmail.com}}

\usepackage{amsmath,amssymb,amsthm,mathrsfs,amsfonts,dsfont}

\def\beq{\begin{equation}}
\def\eeq{\end{equation}}
\def\bea{\begin{eqnarray}}
\def\eea{\end{eqnarray}}

\newcommand{\Qa}{\mathcal{Q}}

\newcommand{\il}{~}

\def\be{\begin{equation}}
\def\ee{\end{equation}}
\def\bea{\begin{eqnarray}}
\def\eea{\end{eqnarray}}

\graphicspath{{pug/}}

\begin{document}

\begin{abstract}
We present the definition of metric bundles in axially symmetric geometries and give explicit examples for solutions   of Einstein equations.
These structures have been introduced in \cite{remnants}
to explain some   properties of  black holes (\textbf{BHs}) and naked singularities (\textbf{NSs}),
 investigated through the analysis of the  limiting frequencies of  stationary observers, which are at the base of a  Killing horizon definition for these  black hole spacetimes.
In \cite{remnants},
 we introduced the concept of \textbf{NS} Killing throats and bottlenecks associated to, and explained by, the  metric bundles.
In particular, we proved that  the horizon frequency  can point out a connection between \textbf{BHs} and \textbf{NSs}. We detail this definition in general and review some essential properties of metric bundles as  seen in different  frames and exact solutions.\end{abstract}


\section{Introduction}

The aim of this work is to discuss the main properties of the metric bundles for axially symmetric spacetimes, concentrating on some exact solutions. In this particular case,
a metric bundle is a family of spacetimes defined by one characteristic photon (circular) orbital frequency $\omega$  and  characterized by a particular relation
between the metric  parameters. This concept is used to establish a relation between black holes (\textbf{BHs}) and naked singularities (\textbf{NSs}) spacetimes.
In \cite{remnants}, we performed an analysis of the metric bundles corresponding to the equatorial plane of the  Kerr, Reissner-Nordstr\"om and Kerr-Newman geometries.
The off-equatorial case of  the Kerr spacetime is considered in detail in \cite{termo-BH}.

A metric bundle is  represented by a curve on the so-called extended plane \cite{remnants}, which is the entire collection of a parameterized  family of solutions. For concreteness, we now consider the family of Kerr spacetimes.
All the metric bundles are tangent to the horizon curve as represented on the extended plane. Then, the horizon curve emerges as the envelope surface of the set of metric bundles. It turns out that \textbf{WNSs} (weak  naked singularities), for which the spin-mass ratio is close to the value of the extreme \textbf{BH}, are related to a portion of the inner horizon, whereas strong naked singularities \textbf(\textbf{SNSs}) with $a>2M$ are related  to the outer  horizon.  In addition, \textbf{WNSs} are characterized by the presence of Killing bottlenecks, which are defined as ``restrictions'' of the Killing throats that appear in \textbf{WNSs}.  Killing throats or tunnels, in turn, emerge   through the analysis of the radii $r_{s}^{\pm}(\omega,a)$ of light surfaces, which depend on the frequency of the stationary observers $\omega$  and the spin parameter $a$
\cite{observers,remnants}. In the case of \textbf{NS} geometries,  a  Killing  throat
is a connected region in the $r-\omega$ plane, which is bounded by the radii
   $r_{s}^{\pm}(\omega,a)$   and contains all the stationary observers allowed within the
	limiting  frequencies  $]\omega_-, \omega_+[$. In the case of \textbf{BHs}, a Killing  throat is either a disconnected  region in the Kerr spacetime
	or a  region bounded by  non-regular surfaces in the  extreme Kerr \textbf{BH} spacetime.
	   The limiting case of a     Killing bottleneck occurs in  the extreme Kerr spacetime, as seen in the Boyer-Lindquist frame, where the narrowing actually closes on the \textbf{BH} horizons. {Killing throats and bottlenecks    were grouped in \cite{Tanatarov:2016mcs}  in   structures named ``whale  diagrams'' of the Kerr and Kerr-Newman spacetimes--see also
\cite{Mukherjee:2018cbu,Zaslavskii:2018kix}.
Moreover,   Killing bottlenecks, interpreted
in \cite{remnants} as  ``horizons remnants'' and related to metric bundles in
\cite{remnants,termo-BH}, appear also connected with the concept of  pre-horizon regime introduced in \cite{de-Felice1-frirdtforstati,de-Felice-first-Kerr}.
 The pre-horizon was analyzed in \cite{de-Felice-first-Kerr}. It was concluded
that a gyroscope would conserve a memory of the static or
stationary initial state, leading to the gravitational
collapse of a mass distribution \cite{de-Felice3,de-Felice-mass,de-Felice4-overspinning,Chakraborty:2016mhx}.

More in general, metric bundles have interesting properties that allow us to explore in an alternative way some aspects of the geometries that define the bundle, providing an alternative interpretation of Killing horizons (in terms of a set of solutions--the extended plane) and establishing a connection between \textbf{NHs} and \textbf{BHs}, based on the fact that each bundle is tangent to the  horizon curve. Moreover, as we shall see below, metric bundles highlight some properties of the horizons  that could influence the exterior properties of
\textbf{BH} geometries by means of characteristic frequencies.
The metric bundle concept can have significant repercussions in the study of \textbf{BH} physics, in the interpretation of \textbf{NSs} solutions and  in the  horizons  and  \textbf{BH} thermodynamics.

In this work, we present the definition of  metric bundles and discuss their properties in the context of \textbf{BH} thermodynamics.
We analyze the  Kerr, Kerr-Newman and Reissner-Nordstr\"om metric bundles. The explicit  expressions for metric bundles in the  Kerr-de Sitter spacetime are also given. Finally, we present some concluding remarks.

\section{Metric bundles}\label{Sec:M-Bundles}

We start by considering the case of the Kerr spacetime.
The Kerr metric,  in Boyer-Lindquist (BL) coordinates, can be expressed as
\bea
\nonumber
&& ds^2=-\frac{\Delta-a^2 \sin ^2\theta}{\rho^2}dt^2+\frac{\rho^2}{\Delta}dr^2+\rho^2
d\theta^2+\frac{\sin^2\theta\left(\left(a^2+r^2\right)^2-a^2 \Delta \sin^2\theta\right)}{\rho^2}d\phi^2\\\label{alai}&&-2\frac{a M \sin^2(\theta ) \left(a^2-\Delta+r^2\right)}{\rho^2}d\phi dt\ ,
\\
&&
\Delta\equiv r^2-2Mr+a^2,\quad\mbox{and}\quad\rho^2\equiv r^2+a^2\cos^2\theta \ .
\eea
It describes an axisymmetric,   stationary, asymptotically flat spacetime.
The parameter $M\geq0$  is      interpreted as  the mass  of the gravitational source, while  the rotation parameter  $a\equiv J/M $
({spin})
is  the  {specific} angular momentum, and   $J$ is the
{total} angular momentum of the source. This is a stationary and axisymmetric geometry with
  {Killing} fields  $\xi_{t}=\partial_{t} $ and
$\xi_{\phi}=\partial_{\phi} $, respectively.

In this work, we will consider also the  Kerr-Newman (KN) geometry which corresponds to  an electrovacuum  axisymmetric  solution
  with a net electric charge $Q$,  described by
metric  (\ref{alai}) with $\Delta_{KN}\equiv r^2+ a^2+ Q^2-2M r$.
The solution  $a=0$ and $Q\neq0$ constitutes the static case of the spherically symmetric and charged  Reissner-Nordstr\"om
spacetime.
The horizons and the outer and inner static limits for the \textbf{KN} geometry  are, respectively,
\bea\label{Eq:KN.RN.shown}
r_{\mp}=M\mp\sqrt{M^2-(a^2+Q^2)},\quad r_{\epsilon}^{\mp}=M\mp\sqrt{M^2-a^2 \cos ^2\theta-Q^2}\ ,
\eea
which for  $a=0$, $Q=0$,  and   $a=Q=0$  leads to  $(r_\pm,r_{\epsilon}^\pm)$ in the Reissner-Nordstr\"om,   Kerr and Schwarzschild  geometries, respectively.
Note that  the \textbf{KN}  horizons $r_{\pm}$ can be re-parameterized for the total charge $\mathcal{Q}_T$ and its variation with respect to the  parameter $\mathcal{Q}_T$ is exactly the same as for  the corresponding radii  $r_{\pm}$ in the \textbf{RN} or Kerr solution. This aspect  will be significant in the study of the  metric bundles  dependence    from the two charges $(a,Q)$.

    \medskip

\textbf{On the BHs horizons}

For the analysis  of some properties of the horizons, we focus  for simplicity on the case  $Q=0$.
Then,  for the Kerr \textbf{BH} geometry   the horizons and ergospheres radii are given by
$r_\pm = M\pm \sqrt{M^2 -a^2}$ and
$r_\epsilon ^\pm = M\pm \sqrt{M^2-a^2\cos^2\theta}$, respectively.

Metric bundles are defined as the set of metrics that satisfy the condition $\mathcal{L}_{\mathcal{N}}\equiv\mathcal{L}\cdot\mathcal{L}=0$,
where $\mathcal{L}$ is the  Killing vector $\mathcal{L}\equiv \partial_t +\omega \partial_{\phi}$. Solutions could be either
\textbf{BHs} or \textbf{NSs}. The quantity $\omega$ will be called the frequency or the angular velocity of the bundle.
In  \textbf{BH} spacetimes, this Killing vector defines also the  thermodynamic variables and the Killing horizons.

   \medskip
\textbf{On the Killing vector $\mathcal{L}$ and  the condition $\mathcal{L_N}=0$}

The event horizons  of a spinning \textbf{BH}  are   Killing horizons   with respect to  the Killing field
$\mathcal{L}_H\equiv \partial_t +\omega_H \partial_{\phi}$, where  $\omega_H$ is the angular velocity of the horizons,   representing   the \textbf{BH} rigid rotation (the event horizon of a stationary asymptotically flat solution with matter satisfying suitable hyperbolic equations  is a Killing horizon).

The Kerr  horizons are,   therefore,  {null} surfaces, $\mathcal{S}_0$,
whose {null} generators coincide with the orbits of an
one-parameter group of isometries, i.e., in general,    there exists a Killing field $\mathcal{L}$, which is normal to $\mathcal{S}_0$.
In general, a Killing horizon is a  lightlike hypersurface (generated by the flow of a Killing vector),
where the norm of a Killing vector is null.
In the limiting case of the static Schwarzschild spacetime ($a=0$, $Q=0$) or the Reissner Nordstr\"om spacetime ($a=0$, $Q\neq0$),
 the event horizons are  Killing horizons with {respect} to the  Killing vector $\partial_t$.
More precisely, for  static (and spherically symmetric) \textbf{BH} spacetimes, the
event, apparent, and Killing horizons  with respect to the  Killing field   $\xi_t$ coincide.

 The \textbf{BH} event horizon of
stationary  solutions
have  constant surface gravity (which is the content of the zeroth \textbf{BH}  law-area theorem--
the surface gravity is constant on the horizon of stationary black holes
\cite{Chrusciel:2012jk,Wald:1999xu}).
The \textbf{BH}  surface area
is non-decreasing  (second \textbf{BH}  law  establishing the impossibility  to achieve
by a physical
process a \textbf{BH} state with zero surface gravity.)
Moreover, the \textbf{BH} surface gravity, which is a  conformal invariant of the metric,
may be defined as the  {rate} at which the norm  $\mathcal{L}_{\mathcal{N}}$   of  the Killing vector $\mathcal{L}$  vanishes from
outside ($r>r_+$).
 (For a Kerr spacetime, this is $\mathcal{SG}_{Kerr}= (r_+-r_-)/2(r_+^2+a^2)$ and, however,
the surface gravity   re-scales with the conformal Killing vector, i.e. it  is not the same on all generators but,  because of the symmetries,  it is constant along one specific generator).

 In the extreme case, where  $r_{\pm}=M$, the surface gravity  is zero and, consequently, the   temperature  is  $T_H = 0$,  but its   entropy (and therefore the \textbf{BH} area)  is not null\cite{Chrusciel:2012jk,Wald:1999xu,WW}. This fact has   consequences also with respect to the stability
 against Hawking radiation
(a  non-extremal
\textbf{BH} cannot reach an   extremal case in a finite number of steps--third \textbf{BH} law).
The variation of the \textbf{BH}
mass, horizon area and angular momentum, including the surface gravity and angular velocity on the horizon, are related by the
 first law of \textbf{BH} thermodynamics:
$\delta M = (1/8\pi)\kappa \delta A + \omega_H \delta J$.
In here, the term dependent on the \textbf{BH}  angular velocity represents the ``work term'' of the first law, while  the fact that the surface gravity  is  constant on the \textbf{BH} horizon, together with  other considerations, allows us to
associate it with
the concept of temperature.
More precisely, we can  formalize this relation by writing  explicitly the Hawking temperature as
 $T_{H}= {\hbar c\kappa }/{2\pi k_{B}}$, where $k_{B}$,  is the Boltzmann constant and $\kappa$ is the surface gravity. %
Temperature $T= \kappa/(2\pi)$;
entropy $S= A/(4\hbar G)$, where $A $ is the area of the horizon $A = 8\pi mr_+$;
pressure $p= - \omega_h $;
volume $V= G  J/c^2$ ($J = amc^3/G$);
internal energy $U$= GM ($M = c^2m/G $= mass), where $m$ is the mass.

Here we study metric bundles which are defined by the  condition   $\mathcal{L}_{\mathcal{N}}=0$; therefore, it is convenient
to re-express some of the concepts of \textbf{BH} thermodynamics mentioned before in terms of
$\mathcal{L}_{\mathcal{N}}$. Firstly, the norm $\mathcal{L_{N}}\equiv\mathcal{L}^\alpha\mathcal{L}_\alpha$ is  constant on the
\textbf{BH} horizon.
Secondly, the constant $\kappa: \nabla^\alpha\mathcal{L_{N}}=-2\kappa \mathcal{L}^\alpha$,
 evaluated on the \emph{outer} horizon $r_+$, defines the \textbf{BH} surface gravity, i.e.,
$\kappa=$constant on the orbits of $\mathcal{L}$
 (equivalently, it is valid that
  $\mathcal{L}^\beta\nabla_\alpha \mathcal{L}_\beta=-\kappa \mathcal{L}_\alpha$ and  $L_{\mathcal{L}}\kappa=0$, where $L_{\mathcal{L}}$ is the Lie derivative--a non affine geodesic equation).

\medskip

\textbf{Stationary observers and causal structure}

The condition  $\mathcal{L_{N}}=0$ is also related to the definition of  stationary observers.
{Stationary observers} are characterized by a four-velocity of the form
 $
u^\alpha=\gamma\mathcal{L}^{\alpha}$ ($\mathcal{L}^{\alpha}\equiv\xi_t^\alpha+\omega \xi_\phi^\alpha$); thus,
$\gamma^{-2}\equiv-\bar{\kappa}\mathcal{L}_{\mathcal{N}},$ where
 $\gamma$ is a normalization factor.
The spacetime  causal structure of the Kerr geometry can be then   studied by  considering also   stationary  observers
 \cite{malament}:
\emph{timelike} stationary  particles have limiting orbital frequencies, which are the photon orbital frequencies  $\omega_{\pm}$, i.e.,
 solutions to the condition $\mathcal{L_{N}}=0$:
\bea\label{Eq:ex-ce}
\omega_{\pm}\equiv \omega_{Z}\pm\sqrt{\omega_{Z}^2-\omega _*^2},
\quad
\omega _*^2\equiv \frac{g_{tt}}{g_{\phi \phi}}=\frac{g^{tt}}{g^{\phi\phi}},\quad \omega_{Z}\equiv-\frac{g_{\phi t}}{g_{\phi\phi}}.
\eea
Therefore, timelike stationary observers have orbital frequencies (from now on simply called frequencies) in the interval
$\omega\in]\omega_-,\omega_+[ $.
 Thus, frequencies $\omega_{\pm}$ evaluated on $r_{\pm}$  provide the frequencies $\omega_H^{\pm}$ of the Killing horizons.

For completeness, we also derive the frequencies $\omega_H$ of the horizons in the Kerr-Newman case,
\[\omega_H^-=\frac{a M\left(2 M\sqrt{M^2-(a^2+Q^2)}-Q^2+2M^2\right)}{4M^2 a^2+Q^4},\quad \omega_H^+=
\frac{aM}{2M \sqrt{M^2-(a^2+Q^2)}-Q^2+2M^2}.\] The limiting Reissner-Nordstr\"om and Kerr  cases can be obtained by imposing
the conditions $a=0$ and $Q=0$, respectively.

\medskip

\textbf{Metric bundles: definition, structure and characteristic frequencies}

Metric bundles are a set of metric tensors
that can include only \textbf{BHs} \emph{or} \textbf{BHs} \emph{and} \textbf{NSs},
such that  each geometry of the set  has, at  a certain radius $r$,  equal  limiting    photon  frequency
$\omega_b\in \{\omega_+,\omega_-\}$, which is called \emph{characteristic bundle frequency}.
Therefore, metric bundles are solution of the zero-norm condition $\mathcal{L_N}(\omega_b)=0$.

It can be  proved that \emph{all} the metric bundles are tangent to the
horizon curve in the extended plane\footnote{An \emph{extended plane} $\pi^+$ is the set of points $(a/M,\Qa)$,
where $\Qa$ is any quantity that characterizes the spacetime and depends on $a$. In general, the extended plane is an
$(n+1)$-dimensional surface, where $n$ is the number of independent parameters that enter $\Qa$
 \cite{remnants}.}--see Fig.\il(\ref{Fig:NodealBXs}). Then, the horizon curve
emerges as the envelope surface of the set of metric bundles.
As a consequence, in \cite{remnants} we introduced the concept of
weak  naked singularities  (\textbf{WNSs}) as those metrics  related to a portion of
the inner horizon,
whereas strong naked singularities (\textbf{SNSs})
are related to the outer horizon in the extended plane.

It can be proved  that all the frequencies $\omega_{\pm}$, in any point of a \textbf{BH} or \textbf{NS} geometry,
are horizon frequencies in the extended plane or,  in other words, since the metric bundles are tangent to  the horizon curve,
each characteristic frequency of the bundle $\omega_b$ is a horizon frequency $\omega_b=\omega_H^x$,
where  $\omega_H^x\in\{\omega_H^-,\omega_H^+\}$.

For seek of clarity, first we  formalize   this definition for the Kerr case as   a one-parameter family of solutions parameterized with the spin  $a$ (or $a/M$). The generalization to the case of several parameters is straightforward as, for example, in  \textbf{KN} and
$\textbf{RN}$ geometries. These cases will be also addressed explicitly below. Particularly, the frequency $\omega_b$ of the bundle is the inner  or outer  horizon frequencies of the spacetime, which is tangent to the horizon at a radius $r_g$ and a spin  $a_g$
(\emph{bundle tangent  spin}). In addition, the bundle is characterized by the  frequency $\omega_0$ of the \emph{ bundle origin}, i.e.,
the point $r=0$ and $a=a_0$  in the extended plane.
Thus, the metric bundles are all characterized by a frequency  $\omega_b=\omega_H^x(a_g)$, where $a_g$ is the bundle tangent spin,
and the frequency $\omega_0$ at $r=0$, where $a=a_0$. The relation between   $a_0,\ a_g,\ r_g,$ and $\omega_b$,
significant for the bundle characterization,  is particularly  simple in the case of a spherically symmetric geometry or on
the equatorial plane of an axisymmetric geometry. However,  in general, the relation,  involving also the bundle origin $a_0$,  depends on the plane  $\sigma\equiv \sin^2\theta$\cite{termo-BH}.

 Metric bundles can be closed on the horizon. In \cite{remnants},
this property has been shown to be due  to the rotation of the  singularity:
the  curves, which define the \textbf{BH} horizons for the static \textbf{RN} case, can be are open;
the  analysis of the \textbf{KN} case represented in \cite{remnants}
 shows  the influence of the spin in the bending and separation into two  families of  curves on the equatorial plane.
On the other hand, in \cite{termo-BH} we proved that, on planes with $\sigma<1$, there can be open Kerr bundles.
 Then, metric bundles of axisymmetric spacetimes have a non-trivial extension corresponding  to negative bundle  frequencies $\omega_b<0$.  These bundle extensions, associated  to characteristic  frequencies  $\omega_b=-\omega_H^{\pm}$ equal in magnitude to the horizon frequencies, clearly  are not  tangent to the horizon curve in the negative frequencies extension of the extended plane.
However,  these  bundle branches  are tangent to the horizon curve in the plane with positive frequencies $-\omega_b>0$.

\medskip

\textbf{Horizon relations for Kerr geometries on the equatorial plane $\sigma=1$}

\medskip

\textbf{Horizons relations \textbf{I:}} \emph{origin frequencies }$\omega_0^{-1}\equiv a_0^{\pm}/M=\frac{2 r_{\pm}(a_g)}{a_g}\equiv \omega_H^{-1}(a_g)$,  \emph{horizons  frequencies } $\omega_H^+(r_g,a_g)=\omega_0=Ma_0^{-1}$, $\omega_H^-(r^{\prime}_g,a_g)=\omega^{\prime}_0=M/a_0^{\prime}$
where  $r^{\prime}_g\in r_-$ ($r_+=r_g$,  $r_-=r_g^{\prime}$).
\\

\textbf{Horizons relations \textbf{II:}} $\omega^{\prime}_0=\frac{1}{4 \omega_0}$, $ \omega_H^+\omega_H^-=\frac{1}{4}$, ($a_0^+(a_g)a_0^-(a_g)=4M^2$), $a_0^{\pm}/M=\frac{2 r_{\pm}(a_g)}{a_g}$.
See \cite{remnants}.

\medskip

In the Kerr  metric bundles,   the Killing vector  $\mathcal{L_{N}}$ is  a function of $r,\ a$ and $\sigma\equiv\sin ^2\theta$. The equatorial plane  is a notable case, showing  in many aspects   similarities with the case of static limiting geometries,
  where $\mathcal{L_{N}}$ is a function of $r$ and $a$, only.

\medskip

\textbf{Explicit form of the metric bundles}
Here, we present  explicit expressions for the KN  metric bundles and their limits:
\bea\label{Eq.lcospis}
&&
\textbf{Kerr geometries-equatorial plane $\sigma=1$:}
\\
\nonumber&&
a_{\omega}^{\pm}(r,\omega;M)\equiv\frac{2 M^2 \omega \pm\sqrt{r^2  \omega ^2 \left[M^2-r (r+2M)  \omega ^2\right]}}{(r+2M)  \omega ^2},
\\
&&\label{Eq:Muss.churnc-parl}
\textbf{KN geometries-equatorial plane $\sigma=1$: }
 \\
 &&\nonumber a_{\omega}^{\mp}=\frac{\mp\sqrt{r^4 \omega ^2 \left\{\omega^2 \left[Q^2-r (r+2M)\right]+M^2\right\}}+\omega M (Q^2 -2 r M) }{\omega ^2 \left[Q^2-r (r+2M)\right]},
\\
&&\label{Eq:Q-para-metric}
\mbox{or}\quad
(Q_{\omega}^{\pm})^2\equiv \frac{r \left\{\omega^2 \left[a^2 (r+2M)+r^3\right]-4 a M^2\omega -rM^2+2M^3\right\}}{(a \omega -M)^2} ,
\\
&&
\textbf{RN geometries}:\quad (Q_{\omega}^{\pm})^2= r \left(\frac{r^3}{M^2} \omega ^2-r +2M\right).
\eea
In the \textbf{RN} geometries, the limiting frequencies are $\omega_{\pm} =\pm \frac{M\sqrt{Q^2+(r-2M) r}}{r^2}.$
The \textbf{KN} frequencies $\omega_{\pm}$ do not depend explicitly on  $\mathcal{Q}_T$; this means that the electric and rotational  parameters  of the geometry play a different role in the solutions $\omega_{\pm}$=constant.

Explicitly, if we   consider a surface $a_g(a_0;Q)$ of the tangent bundle spins  in the case $a_0\neq0$,
 where $Q$ is a parameter, we obtain
\bea&&\nonumber
\textbf{KN bundle origin spin-equatorial plane:}\quad
a_0= \frac{2M^2-Q^2\mp2 M\sqrt{M^2-(a^2+Q^2)}}{a}\quad (r_\mp),
 \\&&\label{Eq:gresti}
 \textbf{KN bundle tangent  spin:}\quad
 a^\mp_g(a_0)=\frac{a_0 \left(2M^2-Q^2\right)\mp2 M\sqrt{a_0^2 \left(M^2-Q^2\right)-Q^4}}{a_0^2+4M^2} %
\\
 &&\mbox{where }\quad a_0>a_L(Q)\equiv\sqrt{-\frac{Q^4}{Q^2-M^2}}\quad\mbox{with}\quad
Q^2\in]0,M^2[\ .
\eea
These functions are very important to derive a relation between \textbf{BHs} (with tangent spins $a_g$) and \textbf{NSs}
(with origin spins $a_0$), as discussed in \cite{remnants}, and also the transformation laws    for
\textbf{BHs} in the  extended plane, as explicitly shown in \cite{termo-BH}.

The relation between \textbf{BHs} and \textbf{NSs} can be  formalized  by analyzing the function of the
tangent spin $a_g(a_0)$   in terms of the bundle origin $a_0$ as follows
\bea\label{Eq:agar}&&
\textbf{Kerr geometry $\sigma=1$}\quad\forall \; a_0>0,\quad a_g\equiv\frac{4 a_0M^2}{a_0^2+4M^2}\quad\mbox{where}\quad a_g\in[0,M]\\
&&\nonumber \mbox{and} \quad \lim_{a_0\rightarrow0}a_g=
\lim_{a_0\rightarrow\infty}a_g=0,\quad a_g(a_0=2M)=M.
\eea
Alternatively, we can explicitly write the relation between the tangent spin and the radius as follows:
\bea\label{Eq:a-tangent}
&&
a_{tangent} (r)\equiv \frac{r (M-r_g)+M r_g}{\sqrt{-(r_g-2M) r_g}}\\&&\nonumber \mbox{where}\quad r_g\in[0,2M],
\quad a_g=a_{\pm}:\quad \frac{r_g}{M}\equiv\frac{2 a_0^2}{a_0^2+4M^2}.
 \eea

\textbf{Some general results from the study of metric bundles in the extended plane}

We now summarize some general results obtained in  \cite{remnants,termo-BH}. For simplicity, we focus on the equatorial plane of the Kerr geometry so that a metric bundle can be represented as a curve on the plane  $(a,r)$.

\medskip

\textbf{Vertical lines $r=$constant in the extended plane}

Vertical lines  $r=$constant in the extended plane intersect  specific metric bundles. First, on a point $r$,
there is always a maximum of two intersections (limiting cases are on the horizon curve or on the origin $r=0$ and $a_0=0$ or
$r=2M$ and $a_0=0$),
which provide the two limit frequencies  $\omega_{\pm}\equiv \{\omega_b,\omega_b^{'}\}$, corresponding to the two
characteristic frequencies of the two metric bundles.
These are also   horizon frequencies $\omega_{\pm}\equiv \{\omega_b,\omega_b^{'}\}\equiv \{\omega_H^x(a_g),\omega_H^y(a_g^{'})\}$, respectively, where  $(x,y)=\pm$ and $a_g$ and $a_g^{'}$. They are the tangent spins of the two bundles with frequency  $\omega_b$ and
$\omega_b^{'}$, respectively.
 We clarify in  \cite{termo-BH} the precise correspondence between  $\{x,y,\pm\}$. In fact, these quantities
 are related to the notion of \textbf{BH} inner horizon confinement,  discussed firstly in \cite{remnants}, and to the horizon replicas introduced in \cite{termo-BH}.
The \textbf{BH} inner horizon confinement is related to the notion of bottleneck as well. It is based on the fact that it is not possible
to find a bundle outside the outer  event horizon ($r>r_+$)  in the plane (and for any geometry $a$) with a characteristic frequency equal to that of the inner  horizon. This implies that outside the horizon of a given spacetime, it is not possible to find a photon limiting  frequency equal to the inner horizon frequency. Nevertheless, it is possible to find such orbits
for the frequencies of the outer horizon.
However, it is possible to find  frequencies of the inner horizon in the Kerr case for  $\sigma$ sufficiently small
(sufficiently close to the rotation axis); therefore, it is possible to "extract" this inner horizon frequency on an "orbit"
$r>r_+:\mathcal{L}\cdot\mathcal{L}=0$.

This notion led to the definition in \cite{termo-BH}
 of the  horizon replicas. These structures occur when there is a point  $r$ of the bundle such that the
  characteristic bundle frequencies $\omega_b(a)\in\{\omega_H^+(a_p),\omega_H^-(a_p)\}$ are located exactly at $r_\pm(a_p)>r_+(a)$,
	that is, on  the horizon with frequency $\omega_b(a)$. Such orbits are, therefore, called horizons replicas
	(these are clearly related to the vertical lines of the extended plane crossing the horizon curve on the tangent point to the bundle).

  \medskip

  \textbf{Horizontal lines $a=$constant on the extended plane}

  Horizontal lines $a=$constant on the extended plane determine a particular geometry and are related to the orbits with frequencies
	equal to that of the Killing horizons in the extended plane  and, therefore,  to the concept of  horizon replicas.

  \medskip

  \textbf{The Kerr-de-Sitter metric bundle}

  To complete this  overview of the metric bundles of axisymmetric  spacetimes, we present  here the explicit expressions for
the Kerr-de Sitter geometry, which has an interesting and complex horizon structure. Further details on these specific solutions can be found in \cite{next}.

 {\small{
 \bea&&\label{Eq:KdeS}
 \textbf{Kerr-de-Sitter metric bundle, general form in $\Lambda$:}
 \\\nonumber
 &&
\Lambda_{\omega}\equiv \frac{6 \left[-\sin ^2(\theta ) \left[\omega ^2 \left(a^2+r^2\right)^2+a^2-4 a M r \omega \right]+a^2 \omega ^2 \sin ^4(\theta ) \left[a^2+r (r-2 M)\right]+a^2+r (r-2 M)\right]}{\left[a^2 \cos (2 \theta )+a^2+2 r^2\right] \left[a \sin ^2(\theta ) \left[a \omega ^2 \left(a^2+r^2\right)-2 \omega  \left(a^2+r^2\right)+a\right]+r^2\right]}\ .
 \eea}
 This expression gives the form of the metric bundles in the Kerr-de-Sitter spacetime in terms of the cosmological constant $\Lambda>0$
for any plane  $\sigma\equiv\sigma^2\theta$.
 Similar solutions can easily be found in terms of  $a_{\omega}$. The extended plane is represented, however, a 3D space.
 In Figs.\il(\ref{Fig:NodealBXs}), we show different representations of this case.

\begin{figure}
  \begin{center}
  \includegraphics[width=13cm]{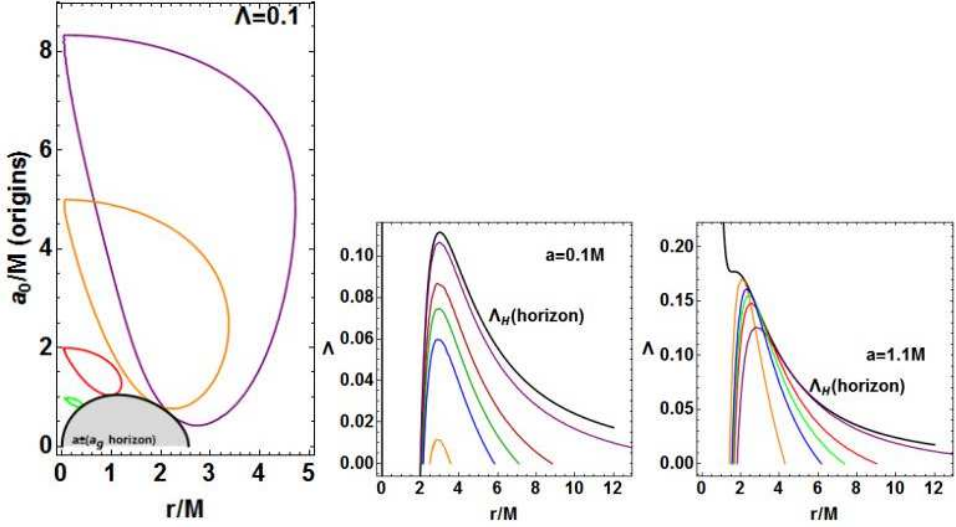}
  \end{center}
  \caption{Kerr-de-Sitter geometry: Equatorial plane ($\sigma=1$). Left panel: Metric bundles of the Kerr-de-Sitter geometries
	in the plane $(a/M, r/M)$ for fixed cosmological constant $\Lambda>0$. The black thick curve is the horizon curve in the extended plane. Metric bundles are tangent to the horizon curve. The origin spins $a_0$ are also shown. The tangent spin  $a_g$ is on the horizon curve. Each bundle (curve) has a specific frequency (the lower bundle corresponds to greater frequencies due to the fact that the inner horizon frequency  is always greater then the outer horizon frequency,  a part in the extreme \textbf{BH} case), which is the horizon frequency of the point $(a,r)$ of the bundle, particularly, at the origin $(a=a_0,r=0)$ and tangent point $(a=a_g,r=r_g)$.  Center and right panel: Bundles in the $(\Lambda,r/M)$ plane for different spins and frequencies. Black curves represent the horizon. Each curve is for a different fixed frequency $\omega$ (the lower the curve, the  greater the frequency).}\label{Fig:NodealBXs}
\end{figure}

\section{Concluding remarks}

We discussed the concept of metric bundles of axially symmetric spacetimes.
In Eqs.\il(\ref{Eq.lcospis}), (\ref{Eq:Muss.churnc-parl}) and (\ref{Eq:Q-para-metric})  explicit expression of these bundles are given on the equatorial  plane of the Kerr geometries, Kerr-Newman spacetimes and for the spherically symmetric Reissner-Nordstr\"om spacetime.
In Eq.\il(\ref{Eq:KdeS}),  we present the expression for the  Kerr-de-Sitter geometry.
Figs\il(\ref{Fig:NodealBXs}) illustrate these bundles and their main features such as the  origins  $a_0$ and the tangent points  $a_g$ on the horizon curve in the extended plane, where the metric bundles are represented as curves.
At the end of Sec.\il(\ref{Sec:M-Bundles}), we discussed some results concerning the general properties of the geometries defined by the bundles, as extracted from the analysis of these structures,  such as the \textbf{BH} horizon confinement and horizon replicas.
The issues discussed in this article refer to the study of \cite{remnants},
where the concept of metric bundle was first introduced and the definition of Killing throat and bottleneck for the   Kerr, Kerr-Newman and  Reissner-Nordstr\"om spacetimes were considered. In \cite{termo-BH}, we present the general definition on an arbitrary  plane of the  Kerr geometry and give definition of horizon replicas.

In a future work, we intend to generalize this study to other spacetimes \cite{next} and investigate in detail the consequences for
 the \textbf{BH}  thermodynamical properties as described in Sec.\il(\ref{Sec:M-Bundles}).
\ack

D.P. acknowledges partial support from the Junior GACR grant of the Czech Science Foundation No:16-03564Y.


\def\prc{Phys. Rev. C }
\def\pre{Phys. Rev. E }
\def\prd{Phys. Rev. D }
\def\jcap{Journal of Cosmology and Astroparticle Physics }
\def\apss{Astrophysics and Space Science }
\def\mnras{Monthly Notices of the Royal Astronomical Society }
\def\apj{The Astrophysical Journal }
\def\aap{Astronomy and Astrophysics }
\def\actaa{Acta Astronomica }
\def\pasj{Publications of the Astronomical Society of Japan }
\def\apjl{Astrophysical Journal Letters }
\def\pasa{Publications Astronomical Society of Australia }
\def\nat{Nature }
\def\physrep{Physics Reports }
\def\araa{Annual Review of Astronomy and Astrophysics}
\def\apjs{The Astrophysical Journal Supplement}
\def\na{New Astronomy}

\def\mdash{---}

\bibliography{pug-K}

\end{document}